\documentclass[manuscript]{aastex}

\shorttitle{He~{\sc ii} sources in NGC 1569}
\shortauthors{Buckalew et al.}

\begin{document}

\title{The Starburst-Interstellar Medium Interaction in NGC 1569 \\
I. Location and Nature of He~{\sc ii} Sources Using \\ {\it Hubble Space 
Telescope WFPC2} Imagery\altaffilmark{1}}

\author{Brent A. Buckalew and Reginald J. Dufour}
\affil{Rice University, Department of Space Physics \& Astronomy, \\
Houston, TX  77025, USA\\
Email:  {\tt brent@regultra.rice.edu}, {\tt rjd@rice.edu}}
\and
\author{Patrick L. Shopbell}
\affil{California Institute of Technology, Department of Astronomy, MC 105-24\\
Pasadena, CA  91125, USA \\
Email:  {\tt pls@astro.caltech.edu}}

\and
\author{Donald K. Walter}
\affil{South Carolina State University, Department of Physical Sciences, \\
Orangeburg, SC  29117, USA \\
Email:  {\tt dkw@physics.scsu.edu}}

\altaffiltext{1}{Based on observations with the NASA/ESA {\it Hubble Space Telescope},
obtained at the Space Telescope Science Institute, which is operated by AURA, Inc., under NASA contract
NAS5-26555.}

\begin{abstract}
We present the detection of He~{\sc ii} sources in the Im galaxy NGC 1569
from {\it HST WFPC2} 
imagery.  Out of the fifteen detections, seven were Wolf-Rayet stars, five 
were stellar clusters with associated He~{\sc ii} emission, 
and three were sources of unknown origin.  The detected Wolf-Rayet stars' colors and magnitudes are
similar to Large Magellanic Cloud late-type WN stars.  The
physical origin of the other He~{\sc ii} sources are discussed.  We conclude
that the equivalent of 51$\pm$19 WNL stars have been detected in NGC 1569, 
and we have estimated the total Wolf-Rayet population in NGC 1569 as 78$\pm$51.  These numbers
are compared with the Wolf-Rayet stellar populations in the SMC, LMC, Milky Way, 
and other starburst galaxies 
relative to their luminosities, dynamical masses, and ionized gas masses.

Previous to this study, 
Wolf-Rayet stars in NGC 1569 were not detected using ground-based 
imagery, but were only indicated through longslit spectroscopy.  This 
is the first time the exact locations of the Wolf-Rayet stars 
in this nearby, well-studied, ``post-starburst" galaxy have been determined.  
\end{abstract}


\section{Introduction}
Evidence of large numbers of 
Wolf-Rayet (WR) stars in galaxies serves as a tracer of recent starbursts. 
The locations of WR stars indicate where star formation occurred 
$\lesssim$ 10 Myr ago and where OB 
stars of mass $\gtrsim$ 20 M$_{\sun}$ evolved to this state
(e.g., Schaerer et al. 1999 and references therein). From the numbers and 
locations of such stars, one can determine 
the starburst properties and can study the kinematical and morphological 
impact the starburst has on the interstellar medium (ISM). Such studies are 
particularly important  
in dwarf irregular galaxies
because of their low escape velocity ($\sim100~{\rm km~s}^{-1}$;
Drissen, Roy, \& Moffat 1993).  Because starburst galaxies have large 
star formation rates (e.g., log $\Sigma_{{\rm SFR}} \approx -0.8$ M$_{\sun}$ yr$^{-1}$ kpc$^{-2}$; 
Kennicutt 1998), the 
number of massive stars is enhanced.  With the 
detection of massive WR 
stars in these disturbed systems, it is possible to correlate the 
morphology of the gas (both H~{\sc i} and H~{\sc ii}) and the recent star formation history.   

NGC 1569 is a nearby (D~=~2.2$\pm$0.6 Mpc; 
Israel 1988) Im galaxy
that has been well-studied over the last 20 years.  
Two prominent, stellar-like features 
are located at the center of this galaxy, which are 
thought to be super-star clusters (SSCs; cf. Prada, Greve, \& McKeith 1994) 
formed in a starburst 2-10 Myr ago \citep{gonzalez97}. It is speculated that they will evolve
into globular clusters. Recent results suggest that SSC A
is two stellar clusters superimposed on one another \citep{demarchi97}.  
Identifying 45 clusters
within NGC 1569, \citet{hunter00} determined their ages which range from 2 Myr to 1 Gyr, and they found 
a subconcentration of clusters, including SSC A, had an age of 4-5 Myr old. 

From multiwavelength studies 
the gaseous morphology of NGC 1569 shows evidence of high 
supernovae activity.  Several studies 
using optical interference-filter imagery of
the ionized gas show evidence of an 
eruptive event which occurred in the galaxy's past \citep{hodge74,vaucouleurs74,waller,hunter93,devost97}.
Kinematics of the ionized gas
that were studied by \citet{tomita94} showed the 
expanding gas moves
at speeds from 10 to 100 km s$^{-1}$.  Heckman et al. (1995) 
 found that the optical filaments at distances of 2 kpc from the 
center of the galaxy are traveling over 200 km~s$^{-1}$.  
In the infrared, \citet{hunter89} observed the dust radiating warmer at 60-155 
$\mu$m, which is explained by the starburst's strong  
radiation field, than dust in similar, irregular galaxies.  
Radio observations also show relics of the eruptive 
starburst.  Israel \& de Bruyn (1988) deduced a high frequency cutoff at 8$\pm$1 GHz, 
which they attributed to a decrease in relativistic electron injection 
about 5 Myr ago.
Observations of \ion{H}{1} \citep{reakes80,israel90,stil98} revealed that 
the distribution of neutral hydrogen is a clumpy ridge or disk surrounded 
by arms which mimic 
the H$\alpha$ arms seen in the optical.  
An \ion{H}{1} hole was detected surrounding the 
SSCs
and was formed after the starburst \citep{israel90}.
The distribution of the global CO 
emission is similar to the \ion{H}{1} distribution 
\citep{young84}.  High resolution CO maps show large (compared to ones in the Galaxy) 
giant molecular clouds surrounding the SSC A \ion{H}{1} hole \citep{taylor99}. 
Finally, ROSAT PSPC and HRI images show an extended 
soft X-ray component perpendicular to the major axis of the 
galaxy \citep{heckman95,stevens98}.  
ASCA images of a hard X-ray source inside NGC 1569 were interpreted as either  
low-mass X-ray binaries or young supernova remnants \citep{della96}.    

The consensus of these studies is that a starburst occurred approximately
10 Myr ago.  This event (of unknown origin) produced SSC A and B.  However, companions found 
around dwarf irregular systems are common (e.g., Taylor et al. 1995). Stil \& Israel (1998) have  
found a 7$\times$10$^6$ M$_{\sun}$ companion 5 kpc from NGC 1569, which makes it a possible  
explanation for what triggered the starburst.  There are large numbers of 
OB and/or WR stars in the two clusters and 
the surrounding field.  Because of the numbers of massive stars and their rapid evolution
(first giving rise to stellar winds and then supernovae), the galaxy underwent a pronounced
kinematical and morphological change, even disruption, during the past several Myr.  
Evidence suggests that the supernova ejected
material will escape the galaxy and will enrich the intergalactic medium. 

Previously, evidence of WR stars in NGC 1569 came primarily from
spectroscopic studies.  Using this method, WR stars were located in 
the ring nebula to the far East in the 
galaxy \citep{drissen93}, in SSC A \citep{gonzalez97}, and elsewhere within NGC 1569 \citep{ho95,martin97}.  
However, \citet{kobulnicky97} found that most of the galaxy's He~{\sc ii}~$\lambda$4686 emission was 
nebular and their slit locations covered some of the regions where WR stars were previously detected. 
Ground-based narrow-band $\lambda4686$ filter 
imagery of NGC 1569 was attempted with little success in finding WR stars (e.g., one stellar knot with 
a light He~{\sc ii} $\lambda4686$ excess; Drissen, Roy, \& Moffat 1993).    

With the improvement of spatial resolution of the Hubble Space Telescope over
ground-based instruments, it is possible to study the morphology of the 
ionized gas with higher detail and locate weak stellar He~{\sc ii} emission-line
sources.  In this paper, new HST {\it WFPC2} F469N filter images  
are presented in an attempt to locate WR 
stellar activity and to confirm the previous WR detections.  
The observations and data reduction are presented in \S2 with the basic results of that analysis given in \S3.  
Discussion of the implications of our findings is given in \S4.  In \S5 a 
summary of our findings and concluding remarks are made.  

\section{Observations and Reduction}

{\it WFPC2} images of NGC 1569 were taken 23 September 1999 for the Cycle 8 program, GO-8133. 
NGC 1569 was oriented in two of the wide-field chips (WF2 \& WF3) of the camera with a nearby $10^{th}$ 
magnitude star placed 
out of the field of view (see Figure \ref{fig1}).  
The effective plate scale is 0$\farcs$0996 pixel$^{-1}$ (1.07 pc pixel$^{-1}$
at the adopted distance of 2.2 Mpc).  The GO-8133 images used here 
were F469N (\ion{He}{2}), F502N ([\ion{O}{3}]), and F547M ($\sim$ Str\"{o}mgren $y$). 
NGC 1569's radial velocity is $-104$~km~s$^{-1}$. The shift of each emission line is $\sim$~2~\AA, and 
therefore, all emission lines were observed very near the center of the filter transmission curve.  
Observational parameters for these data are found in Table \ref{table1}.  

These data were recalibrated using the best reference files 
and the new STScI ``on-the-fly" calibration (OTFC) system.
OTFC images gave the same image statistics over the wide-field chips 
compared to the manually calibrated ones.  
Additionally, we used OTFC processed archival {\it WFPC2} imagery of NGC 1569 from the Cycle 6 programs GO-6111 and GO-6423,
which were primarily broadband images in \ubvr.  Please refer to \citet{greggio98} for more details on the GO-6111 
imagery and to \citet{hunter00} for discussion of the GO-6423 imagery.  Table \ref{table1} 
has observational parameters for these data as well.  

The GO-6111 data were dithered by a few pixels for a subset of four exposures.  These images needed to be aligned with 
respect to one another, which was done using standard IRAF\footnote{IRAF is distributed by the National Optical 
Astronomy Observatories, which are operated by the Association of Universities for Research 
in Astronomy, Inc., under cooperative agreement with the National Science Foundation.} applications.
For all images, cosmic rays were best removed by using the STSDAS package {\it crrej}.  
The cosmic ray removed images 
were then rotated (in the case of the Cycle 6 data) and aligned with respect to the Cycle 8 F547M image.
The F502N and F469N images were continuum subtracted.  
The total 
counts of several bright field stars were determined in the F547M, F502N, and F469N images.  
Average ratios of the emission line to continuum total 
counts for these stars were found, scaled to the F547M image, and the new F547M image was subtracted from 
each interference-filter image.  This process 
was iterated until the field stars were subtracted out, and the residuals were at 
average levels of the background noise.  

Drissen et al. (1999) used a weighted average between F547M and F439W images for 
continuum subtraction of their F469N image of NGC 2403.  
They indicated that this would 
properly subtract red stars and not leave them as ``holes" if only the F547M were  
employed.  
We attempted this method and found that the weighting heavily favored F547M 
($\sim$99\%), and thus, 
only F547M was used. Furthermore, the only F439W (GO-6111) image covers one-half of the galaxy on our field.

Flux calibration of our continuum-subtracted, interference-filter images was done using 
the {\it WFPC2} exposure time calculator found at STScI's home page.  A 
generic flux of $1.0\times10^{-16}$ ergs cm$^{-2}$ s$^{-1}$, an exposure time of 1000 seconds, and the 
redshift velocity of NGC 1569 ($-104$ km s$^{-1}$) were inputted to 
determine the number of object electrons for the F469N and F502N filters.  
The number of electrons sec$^{-1}$ was converted
to DN sec$^{-1}$, and setting the DN sec$^{-1}$ value 
equal to the generic flux of 
$1.0\times10^{-16}$ ergs cm$^{-2}$ s$^{-1}$, we solved for the flux in one DN sec$^{-1}$.  
These conversion numbers are in units of ergs cm$^{-2}$ s$^{-1}$ DN$^{-1}$ and are 
 $1.084\times10^{-14}$ for F502N and $1.847\times10^{-14}$ for F469N.
 
With the calibrated 
He~{\sc ii} image, we determined whether the He~{\sc ii} emission 
was due to a WR star, nebular source, or associated with a stellar cluster.  
All pixels with counts over 3 times the deviation in the background (3$\sigma$) in the F469N image 
were noted as possible detections.    
Each pixel with apparent emission 
was carefully checked in the individual F469N exposures to see whether the location had been 
struck by a cosmic ray, warm pixel or defect which might account for it being high.  There was also the possibility
that the pixel had a blemish in the continuum image which also might account for it being high.  
If a pixel location was in any of these categories, 
it was thrown out.  What remained after this rejection process was considered a ``good" detection. 

Each pixel that met the above criteria was superimposed on the F502N and F555W images.  
We determined from the superposition whether a good detection's location corresponded to a bright, point-like source 
in one or both of the images. 
The detections meeting the 
requirement of being a point source in F502N but no corresponding point-source in F555W were not found in our search.  
If the good detection had a corresponding point source in F555W 
but not in F502N, 
it was labeled S (for stellar sources) or C (for cluster sources).  If the good detection was not associated with 
any point source in either of the images, it was labeled U in the figures and tables below.

For the S and C sources, the He~{\sc ii} flux and {\it WFPC2} B and V magnitudes were calculated.  {\it WFPC2} B 
magnitudes were not measured for some sources because the GO-6111 images did not cover some He~{\sc ii} locations, 
or the corresponding point in the F439W image had low signal-to-noise.  
Our magnitude and search criteria were based on the IRAF procedure {\it apphot}.  The He~{\sc ii}
pixel location of each S source 
was chosen as the starting position for the brightest pixel search in the B and V images.  The search 
radius was restricted to within 2 pixels of the He~{\sc ii} location.  The apertures for the stellar sources were 2 
pixels because most of the He~{\sc ii} pixels were concentrated in the crowded stellar region of the galaxy.  
Background counts were taken from an annulus that was one pixel outside the aperture.  The number of pixel centers 
found within the 2 pixel aperture were counted and the total number of counts was computed.  The He~{\sc ii} flux was 
found by using the same IRAF program and parameters, and the total counts were converted to an absolute 
flux.  The same procedure was used in the case of the C sources, except 
the aperture was set to a larger value.  Looking at the F555W images, one visually 
inspected the maximum 
radius of each cluster, and the average radius was 5 pixels. SSC A was set at 10 pixels, and the 
background annulus was appropriately expanded. 
Only the He~{\sc ii} flux was found for the U sources, although the procedures were the same as for the S sources.  

The absolute emission line fluxes and BV magnitudes presented in Tables \ref{table3} 
through \ref{table6} have not been corrected for reddening.  \citet{devost97} infer line-of-sight 
extinction due to 
the Galaxy in the direction of NGC 1569 
is E(B-V) = 0.50 (A$_V$ = 1.6 mag), and the mean intrinsic extinction of NGC 1569 is E(B-V) = 0.20 (A$_V$ = 0.6 mag). 
 We will adopt these 
values for this study.  Using these numbers along with the extinction curve from Seaton (1979), we found the 
f($\lambda$) value of 0.042 for He~{\sc ii} and the reddening correction of 11.8.  
To correct the V and B magnitudes, a value of 28.9 mag must be subtracted to produce an absolute magnitude. This number was 
computed using the equation
\begin{equation} 
(V-M_{V})_{\rm NGC 1569} = 5\log({\rm D})-5+{\rm A}_{V}
\end{equation}
where D is the adopted distance to NGC 1569, A$_{V}$ is the visual extinction and $V$ is the {\it WFPC2} F555W magnitude 
measured.  Please note that the magnitudes are F555W and F439W magnitudes based on the STScI ``Vega" 
system and are not true Johnson B and V 
magnitudes.  

\section{Imagery Results}

Several candidates were found, which can be characterized into three categories.  
There are seven He~{\sc ii} sources which are associated with a stellar source (hereafter, labeled S), 
five with a cluster (labeled C), 
and three which are found to have none of the above but 
are real He~{\sc ii} sources (labeled U).  Figure \ref{fig2} shows the locations of the 
S, C, and U sources (circles) with respect to the rest of the galaxy.  The underlying gray-scaled image is F555W 
(net exposure time = 930 s) from GO-6423.

Figures \ref{fig2a} and \ref{fig2g} show ``postage stamp" enlargements of each individual S source,   
Figure \ref{fig3a} shows each C source, and   
the U sources are found in Figure \ref{fig5a}.    
The gray-scaled image is the F555W image and the contours are He~{\sc ii} for that particular source.  Each
F555W image has been smoothed using the IRAF command {\it magnify}.  This has made the images more presentable.  However,
the small continuum sources are too blurred in some cases to see.  
The contours are set at 60\%, 40\%, 20\%, and 10\% of the peak 
He~{\sc ii} flux in all C sources excluding C3 and are set at 100\%, 80\%, 60\%, and 40\% in C3 and all S and U sources. 
All figures are 3$\farcs$1 by 3$\farcs$1, and the orientation of each image is the same as 
Figure \ref{fig1}.  

\section{Discussion}
\subsection{Global Properties of Sources}
We now compare our absolute magnitudes and colors of the NGC 1569 WR stars (S sources; Table \ref{table3}) 
to averaged Large Magellanic Cloud 
(LMC) WR stellar magnitudes and colors 
taken from \citet{feitzinger83}.  The comparison with LMC stars is a logical assumption since NGC 1569 is 
classified as Im and their metallicities are similar.  Average S 
absolute visual magnitudes (M$_V$ = -7.43) best agree with average late-type WN magnitudes 
(M$_V$ = -6.26 \& (B-V)$_o$ = -0.07; Feitzinger \& Isserstedt 1983). 
The colors of the S sources ((B-V)$_o$ = -0.2) 
are comparable with early-type WN stars (M$_V$ = -4.39 \& (B-V)$_o$ = -0.22; Feitzinger \& Isserstedt 1983).  
However, the uncertainties due to high reddening and 
to measured magnitudes allow for the agreement of colors of the LMC WNL WR stars and S sources.  Therefore, we hypothesize 
that these S sources are similar to LMC WNL stars.  Spectroscopy of the individual sources would need to be performed in
order to classify each WR star.
 
The He~{\sc ii} luminosity of each S source was checked to Galactic and LMC WR 
stellar luminosities (2.6$\times10^{35}$ -- 2.9$\times10^{36}$ ergs s$^{-1}$; Drissen et al. 1999 and references therein). 
For reference, the average He~{\sc ii} luminosity for an early-type WN (WNE) WR star is 5.2$\pm$2.7$\times$10$^{35}$ 
ergs s$^{-1}$ 
 and 1.6$\pm$1.5$\times$10$^{36}$for a late-type WN WR star (Schaerer \& Vacca 1998 and references therein).  
We find all S sources to have similar luminosities to the Galactic and LMC WR stars 
with S3 (L(He~{\sc ii}) = 4.76$\times$10$^{35}$ ergs s$^{-1}$) 
at the lower end of the luminosities and S2 (L(He~{\sc ii}) = 2.04$\times$10$^{36}$ ergs s$^{-1}$) at the top.   S3 and S4
have He~{\sc ii} luminosities similar to the average WNE He~{\sc ii} luminosity reported by \citet{schaerer98}, and all 
other S sources are 1$\sigma$ outside of this value. 
\citet{massey98b} find that 30 Doradus has several O3If*
stars which were thought to be WN stars in previous surveys.  These stars also emit He~{\sc ii} and have absolute visual 
magnitudes similar to our S sources.  However, the He~{\sc ii} equivalent width of Of stars is typically around 5-10 
\AA~\citep{nota96}, 
and \citet{crowther98} state a He~{\sc ii} luminosity of 1.7$\times$10$^{35}$ ergs s$^{-1}$ for one O3If*/WN star.  
Schaerer \& Vacca (1998 and references therein) also report a He~{\sc ii} luminosity for Of stars of 2.5$\times$10$^{35}$ 
ergs s$^{-1}$.
This amount is a factor of 2 smaller than the faintest S source in He~{\sc ii}.  Furthermore, the detection of O3 stars
would suggest an age $\sim$ 1 Myr \citep{massey98b}, which contradicts the most recent ages calculated for the clusters
in NGC 1569 (4-6 Myr; Hunter et al. 2000).  Also, our sensitivity limit of three times the standard deviation 
of the mean background gives a minimum He~{\sc ii} luminosity of 2.4$\times$10$^{35}$ ergs s$^{-1}$ which places the 
weakest WR stars and Of stars at our detection threshold.  
We must conclude that the S sources are Wolf-Rayet stars (most likely WN stars), but spectroscopy will have to 
be performed to determine their spectral type.  

From Figure \ref{fig2} we see that the stellar sources are concentrated close to 
SSC A.   This supports the idea that the most recent starbursting 
region is primarily concentrated around the super-star clusters A \& B.  The number of WR stars surrounding SSC A 
is greater
compared to SSC B (5 versus 1).  \citet{gonzalez97} did not find WR signatures in the spectra around 
SSC B and 
Kobulnicky and Skillman (1997; cf. their Figure 1) have spectra over these locations and over most of the 
stellar sources listed here.  They find nebular 
He~{\sc ii} in those spectra only.  This leads us to suspect that some of the point sources and He~{\sc ii} emission 
are coincidental, or 
that the star is photoionizing the interstellar medium to produce He~{\sc ii}.  However, Conti (1999) brings up a valid 
point in that 
some starburst galaxies with the same age do not show WR stars and this could be explained by the fact that continuum 
dilution 
of the background starlight could mask over the broad, faint WR emission lines.  Since these spectra came from 
ground-based 
observatories and the slit used in both papers was wide ($\geq$ 1$\farcs$5), it is possible that continuum dilution 
overwhelmed the He~{\sc ii} emission from these 
WR stars in one or both cases.  Finally, the Wolf-Rayet star spectroscopically 
discovered by Drissen \& Roy (1994) is our S7.

Four of the clusters are also found (see Figure \ref{fig3}) in the largest region of star formation.  
One cluster is on the Eastern outskirts of the large star formation region.  
The four clusters within the starburst show both red stars and WR stars (see Figure \ref{fig7}).
The reason we know that there are red stars comes from Drissen et al. (1999). 
They reported that F469N - F555W would 
produce ``holes" where red stars were, and typically there are ``holes" at the center of each cluster. 
Thus, the red stars are separated from the He~{\sc ii} emission.
This was first reported in De Marchi et al. (1997) as a superposition 
of two clusters which make up SSC A.  
In Figure 1 of their paper, they 
label the West cluster SSC A1 and the East SSC A2 (the orientation of their 
figure and ours is roughly the same; see our C4 in Figure \ref{fig7}d).  
SSC A1 is slightly redder than SSC A2.  In our He~{\sc ii} contours, this splitting is also seen.  We 
find that SSC A2 is where the strongest 
He~{\sc ii} emission is while SSC A1 is centered around 
the ``hole" in the He~{\sc ii} contours.  This detection of red supergiants along with Wolf-Rayet stars has been 
well reported in the past and is consistent with the evolution of massive stars (e.g., Massey 1999).  

The locations of the ``unknown sources" are plotted in Figure \ref{fig4}. The three U sources surround SSC A.  
This region is filled with He~{\sc ii} emission which is similar in 
surface brightness and may be related to stellar wind-shocks, or even supernova(e).  
Therefore, it is likely that these He~{\sc ii} 
emissions are nebular in origin.  
Using the Galactic and LMC He~{\sc ii} luminosities above, we see that the He~{\sc ii} emission of the 
unknown sources is comparable (see our Table \ref{table6}).

\subsection{Notes on Individual Sources}
Below, comments are given on each of the detections that were reported in Tables \ref{table3}
through \ref{table6} and shown in Figures \ref{fig2a} through \ref{fig5c}.  

{\it S1} -- Figure \ref{fig2a}a -- One of the more easily discernible stars detected.  It is easy to see 
that the He~{\sc ii} emission is coincident with a bright 
continuum point source.    

{\it S2} -- Figure \ref{fig2b}b -- This star is to the West of SSC A.  The star is blended with SSC A 
due to the use of the IRAF {\it magnify} command.  However, it is the brightest of the WR stellar candidates.  

{\it S3} -- Figure \ref{fig2c}c -- A star due North of SSC A whose
 He~{\sc ii} emission is the lowest of the WR candidates.  The reason for this low amount of emission might be
 because the internal reddening around this star is larger than the average intrinsic reddening in NGC 1569.

{\it S4} -- Figure \ref{fig2d}d -- Again, another WR star that is close to SSC A and has a low He~{\sc ii} emission 
compared to the other candidates.  The arguments for S3 can be applied here.  

{\it S5} -- Figure \ref{fig2e}a -- A WR star that is Northeast of SSC A.  The He~{\sc ii} emission is coincident with 
the bright point source underneath.    

{\it S6} -- Figure \ref{fig2f}b -- This He~{\sc ii} source is close to the star and is probably associated with it 
given the errors in image shifting and rotating.  It is the only WR star close to SSC B.  
The He~{\sc ii} emission is comparable to Galactic and LMC WR stars.  

{\it S7} -- Figure \ref{fig2g}c -- The WR star found spectroscopically 
by Drissen \& Roy (1994).  It is at the larger, brighter source of 
continuum light in the F555W image.  As was 
stated in their article, the ring nebula was probably the result of the stellar wind from the WR star and its 
progenitor.  There are other 
point sources inside the bubble which could be associated with the nebula as well and therefore, helped to contribute 
to the shell's expansion.

{\it C1} -- Figure \ref{fig3a}a -- A cluster within the large H~{\sc ii} region West of SSC A.  
The cluster has two distinct portions.  The Southern part  
has the associated He~{\sc ii} emission, and the Northern portion contains several red stars.  
  It is impossible to 
 know whether they are associated or coincident.  Recall that KS97 
 have spectra showing nebular He~{\sc ii} over this region.  Therefore, we conclude that 
C1's He~{\sc ii} emission is primarily nebular in origin covering an area of 0.0794 arcsec$^2$, 
but there could be WR stars lost in KS97's spectra due to 
continuum dilution.  However, if we assume that all 
the emission is due to WR stars, the number of WNL equivalent stars (using 1.7$\times$
10$^{36}$ ergs s$^{-1}$ for the He~{\sc ii} 
flux of an average WNL star
from Vacca \& Conti 1992) is three.  
This is cluster 10 in \citet{hunter00} and our M$_{F555W}$ magnitude
agrees with theirs given the uncertainties in both measurements. 

{\it C2} -- Figure \ref{fig3b}b -- A cluster lying Southwest of SSC A.  
This also has the separation between red stars and He~{\sc ii} emission.  
The red stars are found at the center of 
the cluster while the brightest He~{\sc ii} is at its Eastern edge. The 
He~{\sc ii} emission is equivalent to four WNLs and has an area of 0.0893 arcsec$^2$.  Also, the colors of the 
system (and C1) are predominantly influenced by red stars.  
This is cluster 13 in \citet{hunter00} and our M$_{F555W}$ magnitude
agrees with theirs given the uncertainties in both measurements.

{\it C3} --  Figure \ref{fig3c}c -- Another cluster where the strongest continuum source is a ``hole" 
in the He~{\sc ii} image signifying red stars. The peak 
of He~{\sc ii} emission is found to the Northeast of the brightest continuum pixel and has an area of 
0.0496 arcsec$^2$. 
Also, there is a slight rise in the [O~{\sc iii}] flux near the peak 
He~{\sc ii} and gives strong evidence for a nebular origin. KS97's 
spectra also cover this region which 
they find only nebular emission.  However, if one assumes that all the emission is from WR stars and 
using the same average emission of a WNL star, the number
of WNL equivalents is two.  This is not classified as a detected cluster in \citet{hunter00}.

{\it C4} -- Figure \ref{fig3d}d -- SSC A.  Gonz\'{a}lez-Delgado et al. (1997) report the finding of 25-40 WNL 
equivalent stars in SSC A.  However, if one uses our numbers 
and assumes that all the emission is from WR stars, the number of WNL equivalents is $\sim$50.  This number 
is larger than their estimate and
suggests that there is some nebular component (with a maximum area of 0.823 arcsec$^2$) that
 we cannot separate in this study.  However, 
continuum dilution of the broad $\lambda4686$ emission line 
might be a problem in the  Gonz\'{a}lez-Delgado et al. (1997) spectra.  Thus, the number must be somewhere 
between the two numbers, probably nearer their estimate.  
This is cluster A in \citet{hunter00} 
and our M$_{F555W}$ magnitude agrees with theirs given the uncertainties in both measurements.

{\it C5} -- Figure \ref{fig3e}e -- A cluster on the very edge of the starburst and may not be a part of the 
starburst which created SSC A and its surroundings. 
Interestingly, this cluster has no red stellar population within or near it.  It is the only cluster like this and is 
far removed from the other four clusters.  
If one assumes that all the He~{\sc ii} emission is from WR stars, the number of WNL 
equivalent stars is three, similar in number to C1--C3.  However, the area of emission
is much smaller at 0.0298 arcsec$^2$.
This is cluster 39 in \citet{hunter00} and our M$_{F555W}$ magnitude
agrees with theirs given the uncertainties in both measurements.

{\it U1} -- Figure \ref{fig5a}a -- The banana-shaped contour is where the maximum level of emission is.  It is on the 
Southern edge of SSC A and is South of S2.  
The emission is probably nebular in 
origin with an area of 0.00992 arcsec$^2$, but the He~{\sc ii} emission is similar to 
the average He~{\sc ii} emission of Galactic or LMC WR stars. 

{\it U2} -- Figure \ref{fig5b}b -- The cluster to the North (V = 18.1; B-V = 0.63) 
of U2 has a large population of red stars. Thus, this He~{\sc ii} emission might be related
to the group of stars just like the situation in the clusters.  
Therefore, it is possible that one WR star is adjacent to the 
cluster.  However, since no star is seen 
its nebular origin is more realistic (with an area of 0.0198 arcsec$^2$).     

{\it U3} -- Figure \ref{fig5c}c -- The circular point source is surrounded by three bright clusters.  No stellar 
point sources are found between the three.  The emission is most 
probably nebular in origin, but could be a WR star since the He~{\sc ii} emission is comparable to 
the average He~{\sc ii} emission of Galactic or LMC WR stars.

\subsection{On the Origin of Nebular He~{\sc ii} in NGC 1569}

The presence of nebular He~{\sc ii} outside of planetary nebulae is rare
because few thermal sources produce enough photons with energies $>$ 54
eV.  \citet{garnett91} stated three alternative ionizing mechanisms which
could account for this nebular He~{\sc ii}.  Hot stellar ionizing continua is
a definite possibility for this galaxy.  The starburst which produced the WR
stars in SSC A occurred only 2-3 Myr ago \citep{gonzalez97} or 4-5 Myr ago \citep{hunter00}.
\citet{schaerer98} modeled strong nebular He~{\sc ii} due to massive
stars in the early starburst phases.  These models have been subsequently updated \citep{schaerer00}.
Using the I(He~{\sc ii}~4686)/I(H$\beta$) ratios from KS97 (their Table
2; values of 8$\times$10$^{-3}$ and 1.2$\times$10$^{-2}$) and Figure 8 from 
\citet{schaerer00}, we conclude that the burst age must be 3-4 Myr old.  This
is in fair agreement with the younger age of the two-stage burst proposed by
\citet{gonzalez97} and the recently calculated ages of SSC A and surrounding clusters 
\citep{hunter00}.

The second mechanism for the formation of He~{\sc ii} is shock excitation such
that the strength of the nebular He~{\sc ii} depends on the velocity of the
shock: ($V_{\rm shock} \approx 120$ km s$^{-1}$; Garnett et al. 1991).  
If shocks were present, bright [O~{\sc iii}] emission would be at the
same location which may have been seen for C3.  At this time, based on our
limited imagery, shocks cannot be considered a viable source.  
Photoionization by
X-rays is the final, alternative explanation for nebular He~{\sc ii}.  \citet{pakull86}
found that nebular He~{\sc ii} was produced around the black hole candidate
binary LMC X-1.  \citet{della96} state that NGC 1569's hard X-ray spectra is due to
the two bright X-ray point sources located in the ROSAT high resolution imager
data.  These sources could be interpreted as low-mass X-ray binaries or young
supernova remnants.  Future X-ray images would resolve whether these point
sources are X-ray binaries and coincident with some of the He~{\sc ii} sources
stated here.  Chandra observations of this object were taken May 2000,
and we will have to wait and see whether this hypothesis is valid.
 Preliminary findings of 
the Chandra data do not show X-ray binaries coincident with any of our He~{\sc ii} sources \citep{kobulnicky00}.

\subsection{Estimate of the Total Number of Wolf-Rayet Stars in NGC 1569}

It was unlikely that we detected all WR stars within NGC 1569, since
many are ``blurred" into SSC A, and this method principally detects WN stars.
We derived a total number of detected Wolf-Rayet stars from our data to be
51$\pm$19.  This was primarily based on the equivalent number of WNL stars in
the C sources.  For deriving the minimum number of WR stars in NGC 1569, we
assumed that all He~{\sc ii} emission in the C sources was nebular unless
previously identified as due to WR stars.  Then, we took the minimum number of
equivalent WNL stars of SSC A (25; Gonz\'{a}lez-Delgado et al. 1997) and the
individual WR stars found in this paper (7).  For the maximum number of WR
stars in NGC 1569, we assumed that all He~{\sc ii} emission in the C sources was 
due to WR stars.  Because we did not know exactly what type of WR stars produce the
He~{\sc ii} emission, we determined the equivalent number of WNL stars within
each cluster.  The total flux of a C source was converted to a luminosity,
divided by the average He~{\sc ii} luminosity of a WNL star (using
1.7$\times$10$^{36}$ ergs s$^{-1}$ from Vacca \& Conti 1992), and the number
(truncated to an integer) gave the equivalent number of WNL stars for that C
source.  This was discussed in Section 4.2, which gave the equivalent number
of WNL stars for each C source.  Thus, we took the maximum number of
equivalent WNL stars of SSC A (50; this paper), the equivalent number of WNL
stars in the other C sources (12) and the individual WR stars (7) to arrive at
our detected number and range.

If we assume that all the nebular He~{\sc ii} is produced from the ionizing
continua of massive stars, then we can estimate the total number of WR stars
in NGC 1569.  Using the values of I(He~{\sc ii}~4686)/I(H$\beta$) from KS97
along with the updated Figure 9 from \citet{schaerer00}, we can determine the
total number of stars.  In the last section, we discussed a probable age of
the clusters as 3-4 Myr.  Over this age range, the ratio of WNL/Total WR stars
is 0.65$\pm$0.35.  Dividing the number of detected WR stars by this percentage
gives us an indication of the total number of WR stars within NGC 1569.
Within roughly a factor of two, the total number of WR stars in NGC 1569 comes
out to be 78$\pm$51.  The updated figures of \citet{schaerer98} were corrected
for an error in the normalization of the fluxes from WR atmosphere models.
Using the old Figure 9 of \citet{schaerer98}, the ratio of WNL/Total WR stars
was 0.3$\pm$0.3 which gave an estimated total WR number of 160$\pm$60.  This
amount of WR stars is similar to the number of observed WR stars in the LMC.
Whereas, the updated model gives a total number closer to the detected amount.

\subsection{Comparisons with Other Galaxies}

It is useful to compare the estimated numbers of WR stars in NGC 1569 with
those for similar galaxies as a function of galaxy properties such as
luminosity, total mass (dynamical), and ionized gas mass.  This we do in
Figures \ref{fig9}-\ref{fig11}. The data used in the production of these
figures came from several sources.  References where we obtained the number
of WR stars in the various galaxies were the following:
\citet{massey92,vacca92,hucht96,izotov97,massey98,bransford99,breysacher99,
drissen99,hucht99, guseva00}. Absolute magnitudes were readily available in
several papers
\citep{vaucouleurs78,thuan81,bergvall86,thronson90,kobulnicky95,guseva00}
and the NASA/IPAC Extragalactic Database.  However, limited data could only be
found on the total (dynamical) masses \citep{
devaucouleurs60,burns71,cottrell76,loiseau81,thuan81,jackson87,dufour90,
taylor93,kobulnicky95,stil98,wilkinson99} and ionized hydrogen masses
\citep{viallefond86,dufour90,sivan90,thronson90,broek91,waller,vacca92,
walterbros94,kennicutt95,mendez99,guseva00}.  Since there are so many sources and
different combinations of data for the ordinate and abscissa values, we have
not labeled the various data points with different symbols.  Because we have
two values (detected and estimated total) of the number of WR stars in NGC
1569, we have plotted both in Figures \ref{fig9}-\ref{fig11}.  The enlarged,
solid NGC 1569 datum point is the detected number of WR stars, and the
enlarged, hollow NGC 1569 datum point is the estimated total number of WR
stars.

Figure \ref{fig9} plots the number of WR stars detected versus the absolute
blue (or photographic) magnitudes of Wolf-Rayet, starburst, and spiral
galaxies.  The ordinate error bars on the NGC 1569 data points indicate our
estimate of the uncertainty due to complications of the nebular
He~{\sc ii} detected in the galaxy and the estimation procedure for the total
number of WR stars as explained in the previous section.   The ordinate error
bars for a majority of the points comes from using the \citet{guseva00} data
because they find the number of WNL stars with various emission
lines. NGC 2403's ordinate error bars are because \citet{drissen99} report 25-40 
individual WR stars.  The abscissa error bars of a majority of the points
come from using the absolute photographic magnitude (M$_{\rm pg}$) from
\citet{guseva00}.  We are assuming that their M$_{\rm pg}$ is $\pm1$ of
M$_{\rm B}$. From the figure, one finds a well defined linear correlation
between the logarithm of the WR stellar count and absolute blue magnitude for
a majority of the Wolf-Rayet galaxies.  This is not surprising, for one would
expect a correlation between the blue luminosity and the number of WR stars,
since WR stars and their O progenitors dominate the blue luminosity of
star-forming galaxies.  However, both NGC 1569 points are found below the
general trend, which could be due to either {\it (a)} NGC 1569 is a 
post-starburst galaxy for which most of the recently formed massive stars have
already evolved through the WR phase, or {\it (b)} our estimated number of WR
stars in NGC 1569 is too low (i.e., no one has detected WC stars in NGC 1569).
We suspect that {\it (a)} is the more likely reason, and that the SMC might
be a similar post-starburst system since the numbers of WR stars in it are
rather accurately known.  Such might not be the case for estimates of WR stars
in the Milky Way and NGC 2403 spirals, which are galaxies likely to have a
more uniform recent star formation rate than starbursting irregulars.

Figure \ref{fig10} is a plot of the number of WR stars versus total mass of
galaxies, using a more limited number of good data available in the
literature, and Figure \ref{fig9} ordinate error bars are used here for
the correlated data.  This graph, albeit quite sparse on good data for the
galaxy masses, suggests that total galaxy mass is not a good tracer of the
observable number of WR stars in galaxies.  This is not surprising since the
total galaxy mass includes dark matter, which does not relate directly to
recent star formation, and the number of WR stars seen in a galaxy depends
critically on the recent star formation efficiency.  Star formation efficiency
would possibly make a better correlator because several of the small galaxies
plotted here are producing similar numbers of WR stars compared to NGC 1569
and the LMC.  Five galaxies (SMC, IC 10, NGC 2403, Mrk 178, and Mrk 209) all seem to
have a relatively low level of recent star formation efficiency.

Finally, Figure \ref{fig11} is a plot of the number of WR stars versus the
ionized hydrogen (H~{\sc ii}) mass, and the Figure \ref{fig9} ordinate error
bars are again used here for the correlated data.  This graph, though again
sparse on good data for other WR galaxies, shows a positive relation between
the number of Wolf-Rayet stars and the mass of H~{\sc ii} (given such a
conclusion is strongly weighted on the SMC and IC 10). Such is not surprising,
given that the young clusters containing WR stars also contain prodigious
numbers of O-stars which ionize the surrounding gas. The abscissa lower limits
derived from spectral properties in \citep{guseva00} also follow this trend.
Several of the lower limits fall within the good data and could signify that
the majority of the ionized hydrogen mass fell within the slit.  The two lower
limits at the right were probably slit locations over larger Markarian
galaxies.

The comparisons between NGC 1569 and other WR and nearby galaxies shown in 
Figures 7-9 further support the idea
that NGC 1569 is seen to be in a ``post-starburst'' phase, but
has had a major, even eruptive, {\it very recent} starburst.  While the
mass of NGC 1569 is $\sim$30 times less than the LMC, it has approximately
similar numbers of WR stars, similar blue luminosity, and a larger H~{\sc ii}
mass.  By comparison with the SMC and IC 10, which have similar to slightly
higher dynamical masses, they have fewer WR stars compared to NGC 1569.  This,
coupled with the extensive system of ionized filaments seen in NGC 1569
\citep{heckman95}, suggests that the magnitude of the very recent
($\sim$10 Myr) starburst in NGC 1569 was exceptional. Today, the
numbers of WR stars and H~{\sc ii} mass found for NGC 1569 reflect the
relatively recent end of this starburst, compared to other nearby, 
star-forming Im galaxies of comparable mass like the SMC and IC 10.  In this
respect, NGC 1569 still retains the properties of starbursting blue-compact
dwarf galaxies compared to typical irregular (Im) systems.  Possibly in
another 50 Myr or so, NGC 1569 will have properties more similar to the SMC,
IC 10, and other irregular systems of similar mass.

\section{Summary}

Using {\it HST WFPC2} imagery, we find the following on the morphology and location of 
He~{\sc ii} sources in the post-starburst galaxy NGC 1569.  

1. Seven He~{\sc ii} sources are associated with a star, indicating that they are WR stars. The 
four WR stars with both magnitudes and colors are consistent with LMC WN stars.
 Spectroscopy of 
the individual WR stars will be performed to confirm their spectral class. 
  
2.  Five clusters have associated He~{\sc ii}. All clusters (excluding SSC A) have emission which would be due to 
2-4 WNL equivalent WR stars.  SSC A, which was discovered by 
Gonz\'{a}lez-Delgado et al. (1997) to have 25-40 WNL equivalent WR stars, has about 50 WNL 
equivalents according to our study.

3.  De Marchi et al. (1997) showed that SSC A was a superposition of two separate clusters.  
We have shown the redder of their two clusters (SSC A1) is the location of a higher concentration of red supergiants, 
and the WR stars are more concentrated in SSC A2.  The coexistence of these two populations is consistent with
massive stellar evolution.   

4.  Three sources close to SSC A were unidentified. The interpretation for these 
sources is that the emission is nebular in origin and possibly related 
to SSC A.  However, all three have He~{\sc ii} emission comparable to the WR stars detected in this survey.

5.  The total detected number of WR stars in NGC 1569 is 51$\pm$19.
The estimated total number of WR stars in 
NGC 1569 is 78$\pm$51. 

If some of these sources are nebular in origin, they could be attributed to either hot stellar ionizing continua 
(which requires the presence of WR stars in most models) or photoionization due to X-rays.  Presently, preliminary 
analysis of the Chandra 
observations of this object do not show any of our sources near X-ray sources. 
However, the model of \citet{schaerer98} and data from KS97 derive similar
starburst ages to those from previous and recent stellar models.  Therefore, a majority of the nebular He~{\sc ii}
is associated with the ionizing continua of the massive stars.  

This paper is the first in a series of papers on the morphological interaction between the starburst and  
interstellar medium of NGC 1569.  
Using other {\it WFPC2} interference-filter images, 
we will 
examine next the detailed morphology of the 
ionized gas components.  The excitation mechanisms of the ionized gas will be modeled both empirically 
and theoretically, and compared with the neutral and molecular gas distribution. 

\acknowledgements
We would like to thank the referee for his/her insightful comments and suggestions.  
We would also like to thank Henry Kobulnicky for allowing us to look at his thesis spectra of NGC 1569. 
B.B.'s research is supported by a NASA-GSRP fellowship (NGT 2-52252).  Obtaining
the HST observations and funding for R.J.D., P.L.S., and D.K.W. via grants related to the Cycle 8 General 
Observer Program GO-8133 were supported by AURA/STScI.

\clearpage

\begin{figure}
\figcaption{[O~{\sc iii}] image of NGC 1569 taken 23 September 1999 for GO-8133.  HST was oriented so that NGC 1569 
fell onto the WF2 and WF3 wide-field chips which left a 10$^{th}$ magnitude star outside the field of view.  
The orientation of all figures throughout this paper will
be the same as this Figure. The field of view is 148$\arcsec~\times~76\arcsec$ (1.579 $\times$ 0.811 kpc). \label{fig1}}
\end{figure}
 
\begin{figure}
\figcaption{Visual continuum images of NGC 1569 taken for GO-6423 with the locations of our He~{\sc ii} sources encircled. 
(a) The Wolf-Rayet candidates (S sources) are identified 
and labeled as they appear in Table \ref{table3}. (b) The clusters with associated He~{\sc ii} emission 
(C sources) are identified 
and labeled as they appear in Table \ref{table4}.   (c) The He~{\sc ii} with neither associated 
[O~{\sc iii}] nor visual continuum (U sources) are identified 
and labeled as they appear in Table \ref{table6}. 
All images in this Figure are oriented and scaled 
as indicated in part (a).
\label{fig2} \label{fig3} \label{fig4}}
\end{figure}

\begin{figure} 
\figcaption{``Postage stamp" enlargements of the fields of S objects 1 through 4.  The size
of each image is 3$\farcs$1$\times3\farcs$1.  The gray-scaled image is F555W and
the contours are of the F469N image of the same area.  The contours are plotted
as the 100\%, 80\%, 60\%, and 40\% level of the peak He~{\sc ii} emission for 
that object.  
\label{fig2a} \label{fig2b} \label{fig2c} \label{fig2d} \label{fig5}}
\end{figure}

\begin{figure} 
\figcaption{``Postage stamp" enlargements of the fields of S objects 5 through 7.  The size
of each image is 3$\farcs$1$\times3\farcs$1.  The gray-scaled image is F555W and
the contours are of the F469N image of the same area.  The contours are plotted
as the 100\%, 80\%, 60\%, and 40\% level of the peak He~{\sc ii} emission for 
that object.  
\label{fig2e} \label{fig2f} \label{fig2g} \label{fig6}}
\end{figure}

\begin{figure} 
\epsscale{0.8}
\figcaption{``Postage stamp" enlargements of the fields of all C objects.  The size
of each image is 3$\farcs$1$\times3\farcs$1.  The gray-scaled image is F555W and
the contours are of the F469N image of the same area.  The contours for all C objects 
excluding C3 are plotted
as the 60\%, 40\%, 20\%, and 10\% level of the peak He~{\sc ii} emission for 
that object.  C3 has contours at the 100\%, 80\%, 60\%, and 40\% level of the peak He~{\sc ii} emission. 
\label{fig3a} \label{fig3b} \label{fig3c} \label{fig3d} \label{fig3e} \label{fig7}}
\end{figure}

\begin{figure} 
\figcaption{``Postage stamp" enlargements of the fields of all U objects.  The size
of each image is 3$\farcs$1$\times3\farcs$1.  The gray-scaled image is F555W and
the contours are of the F469N image of the same area.  The contours are plotted
as the 100\%, 80\%, 60\%, and 40\% level of the peak He~{\sc ii} emission for 
that object.  
\label{fig5a} \label{fig5b} \label{fig5c} \label{fig8}}
\end{figure}

\begin{figure}
\figcaption{Log-Log plot of the number of Wolf-Rayet stars versus the absolute blue magnitude. 
Wolf-Rayet Numbers for the other galaxies come from the following
references: SMC \citep{hucht96}, 
LMC \citep{breysacher99}, MWG \citep{hucht99}, IC 10 \citep{massey92}, 
NGC 2403 \citep{drissen99} and others \citep{vacca92,guseva00}. 
Absolute blue magnitudes were taken from the NASA/IPAC Extragalactic Database (NED), \citet{vaucouleurs78}, 
\citet{thuan81}, \citet{bergvall86}, \citet{thronson90}, \citet{kobulnicky95}, and \citet{guseva00}. The enlarged, solid 
NGC 1569 datum point is the detected number of Wolf-Rayet stars. The enlarged, hollow NGC 1569 datum point 
is the estimated number of Wolf-Rayet stars. \label{fig9}}
 \end{figure}

\begin{figure}
\figcaption{Log-Log plot of the number of Wolf-Rayet stars versus the dynamical mass of the galaxy.
The number of Wolf-Rayet stars for the other galaxies come from the following
references: SMC \citep{hucht96}, LMC \citep{breysacher99}, MWG \citep{hucht99}, IC 10 \citep{massey92}, NGC 2403 
\citep{drissen99}, Mrk 178 \& Mrk 209 \citep{guseva00}, and others \citep{vacca92,izotov97,guseva00}. 
Dynamical masses come from the following sources:
 SMC \citep{loiseau81}, LMC \citep{devaucouleurs60}, NGC 1569 \citep{stil98}, NGC 2403 
\citep{burns71}, MWG \citep{wilkinson99}, IC 10 \citep{cottrell76}, Mrk 178 \& Mrk 209 \citep{thuan81}, 
and others \citep{thuan81,jackson87,dufour90,taylor93,kobulnicky95}.
The enlarged, solid 
NGC 1569 datum point is the detected number of Wolf-Rayet stars. The enlarged, hollow NGC 1569 datum point 
is the estimated number of Wolf-Rayet stars.
\label{fig10}}
\end{figure}

\begin{figure}
\figcaption{Log-Log plot of the number of Wolf-Rayet stars versus the H~{\sc ii} mass of the galaxy.
Square data points are total measurements of the H~{\sc ii} mass. Triangle data points with associated 
arrows are lower limits of the H~{\sc ii} mass computed
using data from \citet{guseva00} and \citet{viallefond86}.
Wolf-Rayet Numbers for the other galaxies come from the following
references: SMC \citep{hucht96}, LMC \citep{breysacher99}, IC 10 \citep{massey92}, NGC 2403 
\citep{drissen99} and others \citep{vacca92,izotov97,massey98,bransford99,guseva00}. H~{\sc ii} masses
were derived using the empirical formula from \citet{kennicutt88}, which 
states that the mass of H~{\sc ii} equals 1.2$\times$10$^5$ L(H$\alpha$)$\times$
10$^{-39}$ M$_{\odot}$. The luminosity of H$\alpha$ came from the following sources: 
SMC \& LMC \citep{kennicutt95}, NGC 1569 \citep{waller}, IC 10 \citep{thronson90}, NGC 2403 
\citep{sivan90}, and others \citep{viallefond86,dufour90,broek91,vacca92,walterbros94,mendez99,guseva00}.
The enlarged, 
solid 
NGC 1569 datum point is the detected number of Wolf-Rayet stars. The enlarged, hollow NGC 1569 datum point 
is the estimated number of Wolf-Rayet stars.
\label{fig11}}
\end{figure}

\clearpage

\begin{deluxetable}{lccccc}
\tablewidth{0pt}
\tablenum{1}
\tablecaption{Observational Parameters of HST/WFPC2 New \& Archival Data \label{table1}}
\tablehead{
\colhead{Filter} &   \colhead{Band/}       & \colhead{PI}      &
\colhead{GO-Program} &
\colhead{Date}          & \colhead{Exposure Time} \\
\colhead{}   & \colhead{Emission Line}      & \colhead{}      &
\colhead{Number} &
\colhead{}          & \colhead{(sec)}}
\startdata
F469N & He~{\sc ii} & Shopbell & 8133 & 23 Sept. 1999 & 2600 \\
F502N & [O~{\sc iii}] & Shopbell & 8133 & 23 Sept. 1999 & 1500 \\
F547M & Str\"{o}mgren $y$ &Shopbell & 8133 & 23 Sept. 1999 & 60 \\
F555W & {\it WFPC2} V & Hunter & 6423 & 21 Oct. 1998 & 930 \\
F439W & {\it WFPC2} B & Leitherer & 6111 & 10 Jan. 1996 & 11280 \\
F555W & {\it WFPC2} V & Leitherer & 6111 & 10 Jan. 1996 & 9720 \\
\enddata
\end{deluxetable}

\clearpage

\begin{deluxetable}{lccccc}
\tablewidth{0pt}
\tablenum{2}
\tablecaption{He~{\sc ii} Point Sources with an Associated F555W Point Source and No [O~{\sc iii}] Point Source \label{table3}}
\tablehead{
\colhead{Number}           & \colhead{R.A.}      &
\colhead{Dec}          & \colhead{F$_{\lambda}$} &
\colhead{V\tablenotemark{b}} &  \colhead{B-V\tablenotemark{c}} \\
\colhead{Number}           & \colhead{(2000)}      &
\colhead{(2000)}          & \colhead{(He~{\sc ii})\tablenotemark{a}} &
\colhead{} &  \colhead{}}
\startdata
S1 & 4~30~48.0 & 64~51~01.9 & 2.26~~(0.2)\tablenotemark{d} & 22.0~~(0.1)  & 0.6~(0.1) \\
S2 & 4~30~48.2 & 64~50~59.0 & 2.98~~(0.2) & 19.35~(0.04) & --        \\
S3 & 4~30~48.4 & 64~50~59.0 & 0.696~(0.2) & 22.8~~(0.5)  & --        \\
S4 & 4~30~48.5 & 64~50~59.2 & 0.975~(0.1) & 21.0~~(0.1)  & 0.5~(0.1) \\
S5 & 4~30~48.6 & 64~50~57.5 & 1.194~(0.1) & 21.3~~(0.1)  & 0.4~(0.1) \\
S6 & 4~30~49.0 & 64~50~53.3 & 2.20~~(0.1) & 21.5~~(0.2)  & 0.5~(0.3) \\
S7 & 4~30~59.2 & 64~50~20.6 & 2.10~~(0.1) & 22.3~~(0.02) & -- \\
\enddata
\tablenotetext{a}{F$_{\lambda}$ measurements are in the units of 10$^{-16}$ ergs sec$^{-1}$ cm$^{-2}$}
\tablenotetext{b}{$(V-M_V)_{\rm NGC 1569} = 5\log(D) - 5 + {\rm A}_V = 26.7\pm0.6 + 2.2\pm0.3 = 28.9\pm0.7$; 
D from Israel (1988) \& A$_V$ from Devost, Roy, \& Drissen (1997)}
\tablenotetext{c}{E(B-V)$_{\rm NGC 1569} = 0.7\pm0.09$ from Devost, Roy, \& Drissen (1997)}
\tablenotetext{d}{Numbers in parentheses are 1$\sigma$ uncertainties.}
\end{deluxetable}

\clearpage

\begin{deluxetable}{lcccccc}
\tablewidth{0pt}
\tablenum{3}
\tablecaption{He~{\sc ii} Sources with an Associated Star Cluster \label{table4}}
\tablehead{
\colhead{Number}           & \colhead{R.A.}      &
\colhead{Dec}          & \colhead{F$_{\lambda}$} &
\colhead{V\tablenotemark{b}}  & \colhead{B-V\tablenotemark{c}} & 
\colhead{Area of He~{\sc ii} Emission} \\

\colhead{}           & \colhead{(2000)}      &
\colhead{(2000)}          & \colhead{(He~{\sc ii})\tablenotemark{a}} &
\colhead{}  & \colhead{} & \colhead{(arcsec$^2$)}}
\startdata
C1 & 4~30~47.3 & 64~51~02.1 & ~~7.21~(0.1)\tablenotemark{d} & 17.41~(0.02) & 1.44~(0.03) & 0.0794 \\
C2 & 4~30~47.5 & 64~50~58.3 & ~~8.81~(0.1)  & 18.80~(0.03) & 1.08~(0.06) & 0.0893 \\
C3 & 4~30~47.7 & 64~51~01.4 & ~~5.89~(0.1)  & 19.39~(0.04) & 0.37~(0.05) & 0.0496 \\
C4 & 4~30~48.2 & 64~50~58.5 & 131.1~~(0.2)  & 15.22~(0.00) & 0.23~(0.00) & 0.823 \\
C5 & 4~30~51.5 & 64~50~48.8 & ~~6.95~(0.07) & 18.91~(0.02) & --   & 0.0298 \\
\enddata
\tablenotetext{a}{F$_{\lambda}$ measurements are in the units of 10$^{-16}$ ergs sec$^{-1}$ cm$^{-2}$}
\tablenotetext{b}{$(V-M_V)_{\rm NGC 1569} = 5\log(D) - 5 + {\rm A}_V = 26.7\pm0.6 + 2.2\pm0.3 = 28.9\pm0.7$;
D from Israel (1988) \& A$_V$ from Devost, Roy, \& Drissen (1997)}
\tablenotetext{c}{E(B-V)$_{\rm NGC 1569} = 0.7\pm0.09$ from Devost, Roy, \& Drissen (1997)}
\tablenotetext{d}{Numbers in parentheses are 1$\sigma$ uncertainties.}
\end{deluxetable}

\clearpage

\begin{deluxetable}{lcccc}
\LARGE
\tablewidth{0pt}
\tablenum{4}
\tablecaption{He~{\sc ii} Sources with Neither [O~{\sc iii}] 
Nor F555W Source \label{table6} }
\tablehead{
\colhead{Number}           & \colhead{R.A.}      &
\colhead{Dec}          & \colhead{F$_{\lambda}$} 
& \colhead{Area of He~{\sc ii} Emission} \\
\colhead{}           & \colhead{(2000)}      &
\colhead{(2000)}          & \colhead{(He~{\sc ii})\tablenotemark{a}} 
& \colhead{(arcsec$^2$)}}
\startdata
U1 & 4~30~48.1 & 64~50~58.5 & 0.861~(0.2)\tablenotemark{b} & $0.00992$ \\
U2 & 4~30~48.3 & 64~50~59.5 & 1.778~(0.2) & $0.0198$ \\
U3 & 4~30~48.4 & 64~51~00.6 & 0.767~(0.1) & $0.00992$ \\
\enddata
\tablenotetext{a}{F$_{\lambda}$ measurements are in the units of 10$^{-16}$ ergs sec$^{-1}$ cm$^{-2}$}
\tablenotetext{b}{Numbers in parentheses are 1$\sigma$ uncertainties.}
\end{deluxetable}


\begin{thebibliography}{DUM}

\bibitem[Bergvall \& Olofsson(1986)]{bergvall86} Bergvall, N. \& Oloffson, K. 1986, \aaps, 64, 469

\bibitem[Breysacher, Azzopardi, \& Testor(1999)]{breysacher99}Breysacher, J., 
Azzopardi, M., \& Testor, G.  1999, \aaps, 137, 117

\bibitem[Bransford et al.(1999)]{bransford99} Bransford, M., Thiker, D., Walterbros, R., \& King, N. 1999, \aj, 118, 1635

\bibitem[van den Broek et al.(1991)]{broek91} van den Broek, A., van Driel, W., de Jong, T., Lub, J., de Grijp, M., \& Goudfrouij, P. 1991, \aaps, 91, 61

\bibitem[Burns \& Roberts(1971)]{burns71} Burns, W. \& Roberts, M. 1971, \apj, 166, 265

\bibitem[Conti(1999)]{Conti99} Conti, P. 1999, IAU Symposium No. 193, {\it Wolf-Rayet Phenomena in Massive Stars and 
Starburst Galaxies}, van der Hucht, K., Koenigsberger, G., \& Eenens, P., eds. pp. 507-516

\bibitem[Cottrell(1976)]{cottrell76} Cottrell, G. 1976, \mnras, 177, 463

\bibitem[Crowther \& Dessart(1998)]{crowther98} Crowther, P. \& Dessart, L. 1998, \mnras, 296, 622

\bibitem[Della Ceca et al.(1996)]{della96} Della Ceca, R., Griffiths, R., Heckman, T., \&
MacKenty, J. 1996, \apj, 469, 662

\bibitem[De Marchi et al.(1997)]{demarchi97} De Marchi, G., Clampin, M., Greggio, L., Leitherer, C., Nota, A., \& Tosi, M. 1997, \apjl, 479, 27


\bibitem[Devost, Roy, \& Drissen(1997)]{devost97} Devost, D., Roy, J., \& Drissen, L. 1997, \apj, 482, 765

\bibitem[Drissen, Roy, \& Moffat(1993)]{drissen93} Drissen, L., Roy, J., \& Moffat, A.
1993, \aj, 106, 1460

\bibitem[Drissen \& Roy(1994)]{drissen94} Drissen, L. \& Roy, J. 1994, 
\pasp, 106, 974

\bibitem[Drissen et al.(1999)]{drissen99} Drissen, L., Roy, J., Moffat, A., \& Shara, M. 1999, \aj, 117, 1249

\bibitem[Dufour \& Hester(1990)]{dufour90} Dufour, R. \& Hester, J. 1990, \apj, 350, 149

\bibitem[Feitzinger \& Isserstedt(1983)]{feitzinger83} Feitzinger, J. \& Isserstedt, J. 1983, \aaps, 51, 505

\bibitem[Garnett et al.(1991)]{garnett91} Garnett, D., Kennicutt, Jr., R., Chu, Y., \& Skillman, E. 
1991, \apj, 373, 458

\bibitem[Gonz\'{a}lez-Delgado et al.(1997)]{gonzalez97} Gonz\'{a}lez-Delgado, R., Leitherer, C., Heckman, T., \& Cervi\~{n}o, M. 1997, \apj, 483, 705 

\bibitem[Greggio et al.(1998)]{greggio98} Greggio, L., Tosi, M., Clampin, M.,
De Marchi, G., Leitherer, C., Nota, A., \& Sirianni, M. 1998, \apj, 504, 725

\bibitem[Guseva et al.(2000)]{guseva00} Guseva, N., Izotov, Y., \& Thuan, T. 2000, \apj, 531, 776

\bibitem[Heckman et al.(1995)]{heckman95} Heckman, T., Dahlem, M., Lehnert, M., Fabbiano, G., Gilmore, D., 
\& Waller, W. 1995, \apj, 448, 98

\bibitem[Ho, Filippenko, \& Sargent(1995)]{ho95} Ho, L., Filippenko, A., \& Sargent, W. 1995,
\apjs, 98, 477

\bibitem[Hodge(1974)]{hodge74} Hodge, P. 1974, \apjl, 191, 21

\bibitem[van der Hucht(1996)]{hucht96} van der Hucht, K. 1996, \apss, 238, 1

\bibitem[van der Hucht(1999)]{hucht99}van der Hucht, K. 1999, IAU Symposium No. 193, {\it Wolf-Rayet
Phenomena in Massive Stars and
Starburst Galaxies}, van der Hucht, K., Koenigsberger, G., \& Eenens, P., eds. pp. 13-20

\bibitem[Hunter et al.(1989)]{hunter89} Hunter, D., Thronson, Jr., H., Casey, S., \& Harper, D. 1989, \apj, 341, 697

\bibitem[Hunter, Hawley, \& Gallagher(1993)]{hunter93} Hunter, D., Hawley, W., \& Gallagher III, J. 1993, \aj, 106, 179

\bibitem[Hunter et al.(2000)]{hunter00} Hunter, D., O'Connell, R., Gallagher, J., \& Smecker-Hane, T. 2000, \aj, in press


\bibitem[Israel(1988)]{israel88} Israel, F. 1988, \aap, 194, 24

\bibitem[Israel \& de Bruyn(1988)]{israel882} Israel, F. \& de Bruyn, A. 1988, \aap, 198, 109

\bibitem[Israel \& van Driel(1990)]{israel90} Israel, F., \& van Driel, W. 1990, \aap, 236, 323

\bibitem[Izotov et al.(1997)]{izotov97} Izotov, Y., Foltz, C., Green, R., Guseva, N., \& Thuan, T. 1997, \apj, 487, 371

\bibitem[Jackson et al.(1987)]{jackson87} Jackson, J., Barrett, A., Armstrong, J., \& Ho, P. 1987, \aj, 93, 531

\bibitem[Kennicutt(1988)]{kennicutt88} Kennicutt, Jr., R. 1988, \apj, 334, 144
\bibitem[Kennicutt et al.(1995)]{kennicutt95} Kennicutt, Jr., R., Bresolin, F., 
Bomans, D., Bothun, G., \& Thompson, I. 1995, \aj, 109, 594

\bibitem[Kennicutt(1998)]{kennicutt98} Kennicutt, Jr., R. 1998, \apj, 498, 541

\bibitem[Kobulnicky et al.(1995)]{kobulnicky95} Kobulnicky, H., Dickey, J., Sargent, A., Hogg, D., \& Conti, P.
1995, \aj, 110, 116

\bibitem[Kobulnicky \& Skillman(1997; hereafter KS97)]{kobulnicky97} Kobulnicky, H., \& Skillman, 
E. 1997, \apj, 489, 636

\bibitem[Kobulnicky(2000)]{kobulnicky00} Kobulnicky, H. 2000, private communication

\bibitem[Loiseau \& Bajaja(1981)]{loiseau81} Loiseau, N. \& Bajaja, E. 1981, RMxAA, 6, 55L

\bibitem[Martin \& Kennicutt(1997)]{martin97} Martin, C., \& Kennicutt, R., Jr. 1997, \apj, 483, 698

\bibitem[Massey et al.(1992)]{massey92} Massey, P., Armandroff, T., \& Conti, P.  1992, \aj, 103, 1159

\bibitem[Massey \& Hunter(1998a)]{massey98b} Massey, P. \& Hunter, D. 1998a, \apj, 493, 180

\bibitem[Massey \& Johnson(1998b)]{massey98} Massey, P. \& Johnson, O. 1998b, \apj, 505, 793

\bibitem[Massey(1999)]{massey99} Massey, P. 1999, IAU Symposium No. 193, {\it Wolf-Rayet Phenomena in Massive Stars and 
Starburst Galaxies}, van der Hucht, K., Koenigsberger, G., \& Eenens, P., eds. pp. 429-438

\bibitem[Mendez et al.(1999)]{mendez99} Mendez, D., Cairos, L., Esteban, C., \& Vilchey, J. 1999, \aj, 117, 1688

\bibitem[Nota et al.(1996)]{nota96} Nota, A., Pasquali, A., Drissen, L., Leitherer, C., Robert, C., Moffat, A. \& 
Schmutz, W. 1996, \apjs, 102, 383

\bibitem[Pakull \& Angebault(1986)]{pakull86} Pakull, M. \& Angebault, L. 1986, \nat, 322, 511

\bibitem[Prada, Greve, \& McKeith(1994)]{prada94} Prada, F., Greve, A., \&
McKeith, C. 1994, \aap, 288, 396

\bibitem[Reakes(1980)]{reakes80} Reakes, M. 1980, \mnras, 192, 297

\bibitem[Schaerer \& Vacca(1998)]{schaerer98} Schaerer, D. \& Vacca, W. 1998, \apj, 497, 618

\bibitem[Schaerer et al.(1999)]{schaerer99} Schaerer, D., Contini, T., \& Pindao, M. 1999, \aaps, 136, 35

\bibitem[Schaerer(2000)]{schaerer00} Schaerer, D. 2000, private communication

\bibitem[Seaton(1979)]{seaton79} Seaton, M. 1979, \mnras, 187, 73P

\bibitem[Sivan et al.(1990)]{sivan90} Sivan, J., Petit, H., Comte, G., \& Maucherat, A. 1990, \aap, 237, 23

\bibitem[Stevens \& Strickland(1998)]{stevens98} Stevens, I. \& Strickland, D. 1998, \mnras, 301, 215

\bibitem[Stil \& Israel(1998)]{stil98} Stil, J., \& Israel, F. 1998, \aap, 337, 64

\bibitem[Taylor et al.(1993)]{taylor93} Taylor, C., Brinks, E., \& Skillman, E. 1993, \aj, 105, 128

\bibitem[Taylor et al.(1995)]{taylor95} Taylor, C., Brinks, E., Grashuis, R., \& Skillman, E. 1995, \apj, 99, 427

\bibitem[Taylor et al.(1999)]{taylor99} Taylor, C., H\"{u}ttemeister, S., Klein, U., \& Greve, A. 1999, \aap, 349, 424

\bibitem[Thuan \& Martin(1981)]{thuan81} Thuan, T. \& Martin, G. 1981, \apj, 247, 823

\bibitem[Thronson et al.(1990)]{thronson90} Thronson, H., Hunter, D., Casey, S., \& Harper, D.  1990, \apj, 355, 94

\bibitem[Tomita, Outa, \& Saitou(1994)]{tomita94} Tomita, A., Outa, K., \& Saitou, M.
1994, \pasj, 46, 335

\bibitem[Vacca \& Conti(1992)]{vacca92} Vacca, W. \& Conti, P. 1992, \apj, 401, 543

\bibitem[de Vaucouleurs(1960)]{devaucouleurs60} de Vaucouleurs, G. 1960, \apj, 137, 265

\bibitem[de Vaucouleurs, de Vaucouleurs, \& Pence(1974)]{vaucouleurs74} de Vaucouleurs, G.,
de Vaucouleurs, A., \& Pence, W. 1974, \apjl, 194, 119

\bibitem[de Vaucouleurs \& Pence(1978)]{vaucouleurs78} de Vaucouleurs, G. \& Pence, W. 1978, \aj, 83, 1163

\bibitem[Viallefond \& Goss(1986)]{viallefond86} Viallefond, F. \& Goss, W. 
1986, \aap, 154, 357

\bibitem[Waller(1991)]{waller} Waller, W. 1991, \apj, 370, 144

\bibitem[Walterbros \& Braun(1994)]{walterbros94} Walterbros, R. \& Braun, R. 1994, \apj, 431, 156

\bibitem[Wilkinson \& Evans(1999)]{wilkinson99} Wilkinson, M. \& Evans, N. 1999, \mnras, 310, 645

\bibitem[Young, Gallagher, \& Hunter(1984)]{young84} Young, J., Gallagher, J., \& Hunter, D.
1984, \apj, 276, 476

\end{thebibliography}
\end{document}